Andra Oltmann*, Roman Kusche and Philipp Rostalski

# Spatial Sensitivity of ECG Electrode Placement

**Abstract:** The electrocardiogram (ECG) is a well-known technique used to diagnose cardiac diseases. To acquire the spatial signal characteristics from the thorax, multiple electrodes are commonly used. Displacements of electrodes affect the signal morphologies and can lead to incorrect diagnoses. For quantitative analysis of these effects we propose the usage of a numerical computer simulation. In order to create a realistic representation of the human thorax including the heart and lung a three-dimensional model with a simplified geometry is used. The electrical excitation of the heart is modelled on a cellular level via the bidomain approach. To numerically solve the differential equations, describing the signal propagation within the body, we use the finite element method in COMSOL Multiphysics®. The spatial gradients of the simulated body potentials are calculated to determine placement sensitivity maps. The simulated results show that the sensitivity is different for each considered point in time of each ECG wave. In general, the impact of displacement is increased as an electrode is more closely located to the signal source. However, in some specific regions associated with differential ECG leads the placement sensitivity distribution deviates from this simple circular pattern. The results provide useful information to enhance the understanding of placed specific effects on classical ECG features. By additional consideration of patient-specific characteristics in the future, the used model has the ability to investigate additional body-related aspects such as geometrical body shape or composition of various tissue types.

______
**\*Corresponding author: Andra Oltmann:** Fraunhofer IMTE, Mönkhofer Weg 239a, Lübeck, Germany, e-mail: andra.oltmann@imte.fraunhofer.de
**Roman Kusche:** Fraunhofer IMTE, Lübeck, Germany
**Philipp Rostalski:** Fraunhofer IMTE, Lübeck, Germany and Institute for Electrical Engineering in Medicine, Universität zu Lübeck, Lübeck, Germany

# 1 Introduction

The electrocardiogram (ECG) is a well-known non-invasive technique used to diagnose cardiac diseases. The electrical excitation of the heart required for the periodic contraction of the heart muscle fibres leads to electrical fields within the body. To acquire the spatial signal characteristics from the human thorax, the potentials can be detected with multiple electrodes on the skin surface. Commonly, a 12-lead ECG is recorded, whereby the electrodes are aligned in a standardised manner to the limbs and thorax. A typical differential ECG signal is characterised by atrial (P wave) and ventricular depolarisation (R wave) as well as by ventricular repolarisation (T wave) [1].

A challenge in signal acquisition is electrode displacements deviating from the ideally used positions. As a consequence, the ECG signals change and incorrect diagnoses can be made. In a study by Bond et al. [2] it was shown that electrode displacements have a probability of 17 - 24 % to influence the signal interpretation. The incorrect placements are caused by variability of physicians in identifying standardised electrode positions [3]. The individual patient anatomy further influences this as the localisation of the heart in the human body is not exactly known [4]. In addition, electrode shifts relatively to the heart induced by body deformation such as lying on one side of the body can be challenging [5].

In general, the electrodes on the chest are more sensitive to displacement than the limb electrodes because they are located closer to the signal source [6]. Therefore, studies such as [2] and [4] investigate the displacements of the precordial leads. They use a large set of electrodes to measure the required body surface potentials which are necessary for displacement analysis on the thorax. The Root Mean Square Error (RSME) is utilized to quantitatively analyse the difference between the signals at the original and shifted electrode position for specific ECG segments [2,4]. In [2] it is measured that the diverse precordial leads influence the segments of the ECG to a different extent if they are misplaced. The results in [4] show furthermore that the influence on signal morphology is enhanced by a larger displacement. Furthermore, for some electrodes it is relevant in which direction a shift is made [4].

As previously described, a quantitative sensitivity analysis of the whole thorax to electrode displacement is often associated with high effort due to the large number of electrode measurements and is therefore not comprehensively investigated in the literature. For this reason and in order to investigate patient-specific aspects in perspective, we propose the usage of heart modelling combined with numerical computer simulation to obtain placement sensitivity maps for the P, R and T wave of the ECG which visualize the influence of electrode displacements.

## 2 Methods

In order to calculate the spatial sensitivity of electrode placement during ECG a numerical computer simulation is performed. To illustrate and demonstrate the possibilities of this approach, a simplified three-dimensional geometry is employed. The model used in this paper is therefore based to the approach of Sovilj et al. [7]. It includes the torso, lungs and the heart with the cardiac chambers and is built up from basic geometric shapes. The heart is divided into seven further subdomains with specific tissue characteristics corresponding to the excitation conduction system. The whole geometry as well as a slice through the heart in the xy-plane is shown in Figure 1 and Figure 2 respectively. The electrical ground (GND) is attached to the lower left corner of the model. The other corners correspond to the standard ECG electrodes ($VL, VR$ and $VR$) used on the limbs.

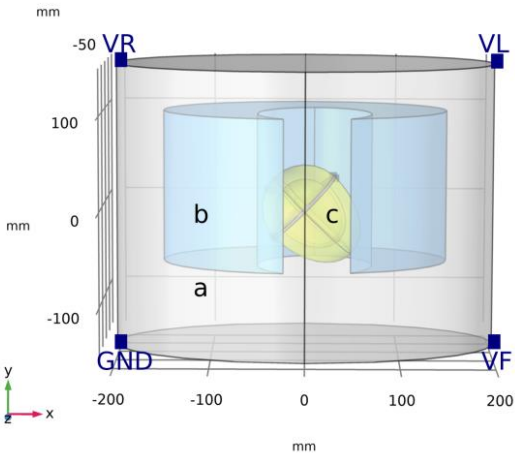

**Figure 1:** Simplified 3D geometry with torso (a), lungs (b) and heart (c) including the cardiac chambers. GND denotes the electrical ground. $VR$, $VL$ and $VF$ are limb electrodes used as standard.

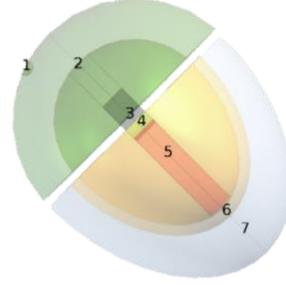

**Figure 2:** Slice through the heart in the xy-plane. The heart is sectioned in sinoatrial node (1), atria (2), atrioventricular node (3), His bundle (4), bundle branches (5), Purkinje fibers (6) and ventricular myocardium (7).

The electrical excitation of the heart is modelled at a cellular level using the bidomain approach based on adjusted FitzHugh-Nagumo equations. These describe an excitable system such as cellular activation in this case [8]. The model governing equations consists of the dependent state variables $V_i$: the intracellular potential, $V_e$: the extracellular potential, and $u$: a variable controlling cellular refractoriness. The intracellular and extracellular electrical conductivities are denoted $\sigma_i$ and $\sigma_e$ respectively. The parameters $a, b, c_1, c_2, d, e, k, A$ and $B$ are region-specific for the heart and modulate the ion current $i_{ion}$ across the cell membrane. All area-related parameters and the initial state values are taken from [7]. Considering that $V_m = V_i - V_e$ is the transmembrane voltage, the following set of equations are defined in each area of the heart [7]:

$$\frac{\partial V_e}{\partial t} - \frac{\partial V_i}{\partial t} + \nabla(-\sigma_e \nabla V_e) = i_{ion} \quad (1)$$

$$\frac{\partial V_i}{\partial t} - \frac{\partial V_e}{\partial t} + \nabla(-\sigma_i \nabla V_i) = -i_{ion}, \quad (2)$$

$$\frac{\partial u}{\partial t} = ke\left[\frac{V_m - B}{A} - du - b\right]. \quad (3)$$

The ion current $i_{ion}$ is formulated according to

$$i_{ion} = kc_1(V_m - B)\left[a - \frac{V_m - B}{A}\right]\left[1 - \frac{V_m - B}{A}\right] + kc_2 u \quad (4)$$

in the sinoatrial node and

$$i_{ion} = kc_1(V_m - B)\left[a - \frac{V_m - B}{A}\right]\left[1 - \frac{V_m - B}{A}\right] + kc_2 u(V_m - B) \quad (5)$$

in all other heart regions respectively. At all inner boundaries of the model which are in contact with the heart, it is specified that there is zero flux for $V_i$. The inward flux for $V_e$ is equal to the current density from the torso and cardiac chambers [7].

To simulate the potentials $V$ in the volume conductors (torso, lung and cardiac chambers) with the electrical conductivities $\sigma_o$ (see Table 1), the following Laplace equation must be solved in these regions [7]:

$$\nabla \cdot (-\sigma_o \nabla V) = 0. \tag{6}$$

**Table 1:** Tissue conductivity values in each domain of the volume conductor [7].

| Domain | Conductivity [S/m] |
|---|---|
| Torso | 0.2 |
| Lung | 0.04 |
| Blood | 0.7 |

All inner boundaries between the heart and the volume conductor are set to $V = V_e$. All exterior boundaries considered electrically isolated [7].

To solve the described set of differential equations COMSOL Multiphysics® v. 5.6 (COMSOL AB, Stockholm, Sweden) is used. The resulting mesh of the applied finite element method consists of 865169 tetrahedral volume elements. The maximum element size varies depending on the domain size between 1 mm and 25 mm. The potentials are simulated for a period of 0.8 s with a temporal resolution of 0.5 ms.

To determine and visualize the spatial sensitivity of the electrode placement, the spatial gradients magnitude of the simulated potentials on the torso surface are calculated. The resulting normalized sensitivity maps are evaluated for points in time of P, R and T wave of the ECG. The Einthoven II lead ($V_{II} = V_F - V_R$) is used to determine these features. The normalized maps were validated by a convergence study with an increasing number of finite elements.

## 3 Results

The simulation results generate a characteristic Einthoven II lead. The signal with the marked evaluated time points of P, R and T wave is shown in Figure 3.

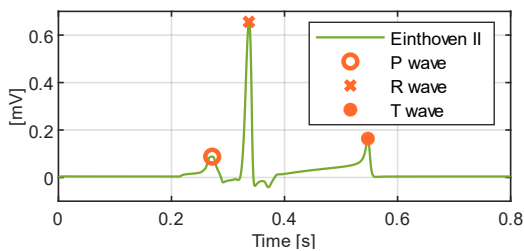

**Figure 3:** Simulated Einthoven II lead with marked evaluated points in time.

As an example, the modelled distribution of the potential $V$ on the torso surface during the R wave is illustrated in Figure 4. The dipole characteristic of the propagating excitation is clearly visible.

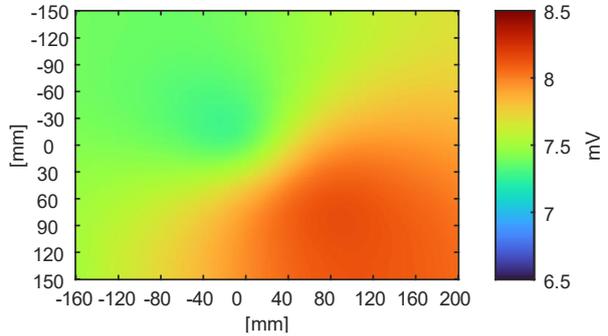

**Figure 4:** Simulated potential distribution on the skin surface during the R wave.

The determined normalized gradient maps demonstrate the spatial sensitivity of electrode placement for the P, R and T wave in Figure 5.

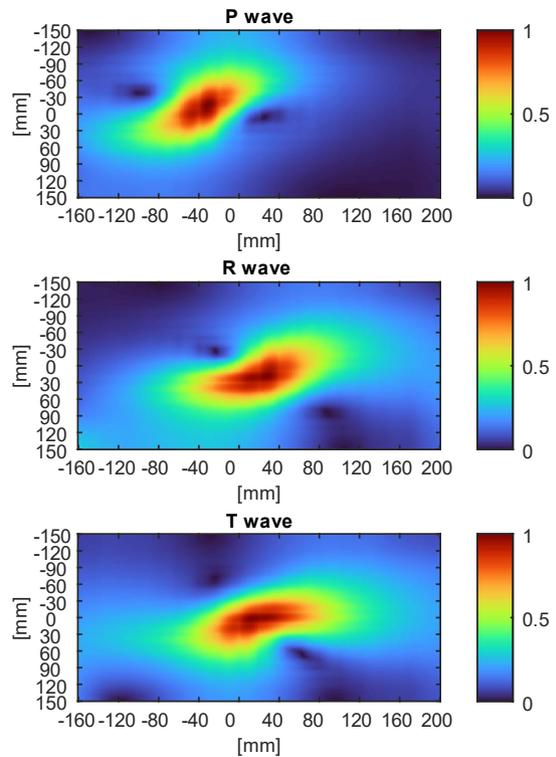

**Figure 5:** Calculated normalized spatial sensitivity maps for the P, R and T wave. Each time point has a different distribution of sensitivity.

The spatial distribution of the sensitive areas is different for each ECG wave and corresponds to the state of electrical excitation in the heart. In general, electrodes are more sensitive to misplacement as they are closer located to the signal source. However, in certain areas where differential ECG electrodes are typically placed, such as the chest wall electrodes, the regional sensitivity differs from this simple circular pattern. Each sensitivity map in Figure 5 shows a kidney-shaped distribution.

## 4 Discussion

The presented simplified three-dimensional heart model, which is solved numerically using the finite element method, visualize electrode displacement effects and provides comparable results with patient measurements as reported in [2] and [4]. In particular, the deviation of the regional sensitivity from the simple circular pattern shows that the influence on the ECG depends on the displacement direction of the electrode.

Currently, the placement maps only show the spatial sensitivity of the signal amplitude at specific points in time of the ECG. Further investigations will be carried out on the simulated output with other diagnostically relevant ECG features such as the analysis of a ST segment which can be an indicator of myocardial infarction [9]. It should be noted that an additional convergence study may be needed for further investigations.

However, the initial results of the presented numerical computer simulation provide a first step for detailed analysis of placed-specific effects on the interpretation of ECG features. In the future, patient-specific characteristics recorded with further biomedical diagnostic techniques can be integrated into the model. This enhancement provides the ability to investigate other body-related aspects such as the geometric body shape or the composition of various tissue types.

**Author Statement**

Research funding: This work was supported by European Union – European Regional Development Fund, the Federal Government and Land Schleswig Holstein, Project: "Diagnose- und Therapieverfahren für die Individualisierte Medizintechnik (IMTE)", Project No. 12420002. Conflict of interest: The last author was supported by Drägerwerk AG & Co. KGaG, Lübeck, Germany. Ethical approval: The conducted research is not related to either human or animals use.